\documentclass[prd,twocolumn,showpacs,superscriptaddress,nofootinbib]{revtex4}

\usepackage{amsfonts}
\usepackage{amsmath}
\usepackage{amssymb}
\usepackage{amsthm}
\usepackage{bm} 
\usepackage{dcolumn}
\usepackage{epsfig}
\usepackage{graphicx}
\usepackage{graphics}
\usepackage[latin1]{inputenc}
\usepackage{latexsym}
\usepackage{rotating}
\usepackage{hyperref}

\newcommand\be{\begin{equation}}
\newcommand\ba{\begin{eqnarray}}
\newcommand\ee{\end{equation}}
\newcommand\ea{\end{eqnarray}}

\newcommand{\pont}{{\,^\ast\!}R\,R}

\begin{document}

\title{Constraining Parity Violation in Gravity \\ 
with Measurements of Neutron-Star Moments of Inertia}


\author{Nicol\'as Yunes}
\affiliation{Department of Physics, Princeton University, Princeton, NJ 08544, USA.}

\author{Dimitrios Psaltis} \author{Feryal \"{O}zel}
\affiliation{Department of Astronomy, University of
Arizona, Tucson, AZ 85721, USA.}
\affiliation{Institute for Theory and Computation (ITC), 
Harvard Smithsonian Center for Astrophysics, 60 Garden St., Cambridge, MA
02138, USA.}


\author{Abraham Loeb} 
\affiliation{Institute for Theory and Computation (ITC), 
Harvard Smithsonian Center for Astrophysics, 60 Garden St., Cambridge, MA
02138, USA.}

\date{\today}


\begin{abstract}

Neutron stars are sensitive laboratories for testing general relativity, especially when considering deviations where velocities are relativistic and gravitational fields are strong.  One such deviation is described by dynamical, Chern-Simons modified gravity, where the Einstein-Hilbert action is modified through the addition of the gravitational parity-violating Pontryagin density coupled to a field. This four-dimensional effective theory arises naturally both in perturbative and non-perturbative string theory, loop quantum gravity, and generic effective field theory expansions. We calculate here Chern-Simons modifications to the properties and gravitational fields of slowly spinning neutron stars. We find that the Chern-Simons correction affects only the gravitomagnetic sector of the metric to leading order, thus introducing modifications to the moment of inertia
but not to the mass-radius relation.  We show that an observational determination of the moment of inertia to an accuracy of $10\%$, as is
expected from near-future observations of the double pulsar, will place a constraint on the Chern-Simons coupling constant of $\xi^{1/4}
\lesssim 5 \; {\rm{km}}$, which is at least three-orders of magnitude stronger than the previous strongest bound.

\end{abstract}


\pacs{04.40.Dg,04.50.Kd,04.60.Rt,04.80.Cc}


\maketitle

\section{Introduction}
\label{intro}

Even though it has been almost a century since its original proposal,
general relativity (GR) remains only marginally tested in the strong,
dynamical regime, where velocities are relativistic and gravitational
fields are strong.  Great effort has gone into testing this theory in
the solar system and with the binary pulsar (via the so-called
parameterized
post-Newtonian~\cite{Nordtvedt:1968qs,1972ApJ...177..757W,1971ApJ...163..611W,1972ApJ...177..775N,1973ApJ...185...31W,lrr-2006-3}
and post-Keplerian
frameworks~\cite{1988grra.conf..315D,1992PhRvD..45.1840D,lrr-2006-3}),
as well as in cosmological settings (e.g.,~via the so-called
parameterized post-Friedmannian
framework~\cite{2002PhRvD..66j3507T,Linder:2007hg,Hu:2007pj,Fang:2008sn}). With
the imminent discovery of gravitational waves, new frameworks have
been proposed (e.g.,~the parameterized post-Einstenian
one~\cite{Yunes:2009ke} or the multipole expansion of black-hole
spacetimes~\cite{Ryan:1997hg,Collins:2004ex,Glampedakis:2005cf,Vigeland:2009pr})
that will allow us to search for GR
deviations with gravitational waves in the strong, dynamical
regime. Until such observations become available, however, one must
rely on the next best tool to test GR in the strong field:
observations of neutron stars and black holes in the electromagnetic
spectrum (see e.g.,~Ref.~\cite{Psaltis:2007rv,lrr-2008-9}).

GR tests with neutron stars appear {\emph{a priori}} to be not as clean as
solar system or binary pulsar tests.  When dealing with weak-field
systems, we can usually employ the effacement 
principle~\cite{Damour:1982wm,Thorne:1984mz} to treat objects as
effective point particles moving under the influence of some force,
which mathematically resembles Maxwell's electromagnetism (the
so-called gravitomagnetic analogy; see eg.~\cite{1988SciAm.258...69P}).  
In modified theories of gravity, however, the effacement principle and Birkhoff's
theorem need not hold, but corrections usually arise at first order in weak-field
perturbation theory~(see e.g.,~\cite{lrr-2006-3} for a review of internal structure effects in
alternative theories).

When dealing with neutron stars, we must inescapably account 
for the matter content of the star properly, as this can greatly influence the astrophysical 
observables to leading order. Our lack of detailed knowledge of their matter content is encoded 
in the equation of state (EOS), which relates the matter density to its pressure. 
Several EOSs have been studied in the literature (see eg.~\cite{1996csnp.book.....G,Lattimer:2000nx}), 
differing mostly in the nature of the strong interaction at very high densities and the presence of a softening
component (such as hyperons, Bose condensates or quark matter), none of which we possess laboratory data for. 
Different possible EOSs thus lead to slightly different electromagnetic observables, which generically
mimic alternative theory effects.  Neutron star tests of GR must thus address
possible degeneracies between the EOS and true GR deviations.

Such a degeneracy between matter content and gravitational theories is
not new to GR tests. For example, the observed acceleration of the
universe can either be attributed to a dark energy fluid or to
modifications to the Einstein equations. This degeneracy in the
explanation of observables is a consequence of the Strong Equivalence
Principle, which essentially states that it is difficult to
differentiate between a gravitational field and a matter distribution,
since they are tied together by Einstein's equations.

The only way to break this degeneracy is to possess at least two
{\emph{independent}} sets of neutron star observations: one to
determine the EOS as a function of the theory under consideration and
another one to pin down the specific theory. Such two-observation
tests have been dramatically successful when testing GR with binary
pulsars, where the degeneracies are associated with the unknown masses
of the pulsars~\cite{lrr-2006-3}. New and exciting observations are
beginning to constrain the neutron star
EOS~\cite{Ozel:2008kb,Ozel:2009da}, suggesting that such
two-observation tests will become possible in the near future.  
In this paper, we consider a particular EOS for neutron-star matter 
that is consistent with current observations. Our results, however, depend  
weakly on our ignorance of the EOS. We show that the corrections introduced by 
the specific alternative theory considered
couples strongly only to large density gradients, which occur only close to the neutron
star surface, where our knowledge of the EOS is fairly robust.

The construction of modified gravity theories is a non-trivial endevour
that can be classified into two groups: infrared modifications and
ultraviolet modifications  (see eg.,~\cite{Yunes:2009ke,Sotiriou:2008rp} for a review).  
The first group deals with corrections to the action that modify the weak-field predictions of GR. 
Examples of these include $f(R)$ gravity~\cite{Carroll:2003wy,Chiba:2003ir,Sotiriou:2006hs}, 
DGP gravity~\cite{Dvali:2000hr}, Horava-Lifshitz gravity~\cite{Horava:2009uw,Lu:2009em}, and
TeVeS~\cite{Bekenstein:2004ne,Bekenstein:2005nv}.  Usually, such
modifications are introduced to propose explanations for the late-time
expansion of the universe or the anomalous rotation curves of
galaxies. The second group corrects the action by introducing higher-order
curvature terms, which, by construction, do not modify the leading-order predictions of GR
in the weak-field. These theories do modify the strong-field regime of gravity, where
neutron star observations can place stringent constraints. Examples include
Chern-Simons (CS) modified gravity~\cite{Alexander:2009tp}, Gauss-Bonnet
modified gravity~\cite{2007CQGra..24..361M}, scalar-tensor
theories~\cite{Damour:1992we,Damour:1993hw,Damour:1996ke}, and certain
$f(R)$ models~\cite{Upadhye:2009kt,Babichev:2009td,Babichev:2009fi,Cooney:2009rr}.

We concentrate here on tests of dynamical CS modified
gravity, which is currently only weakly constrained by binary pulsar
observations~\cite{Yunes:2009hc}. This four-dimensional theory adds a certain contraction
of two Riemann tensors and the Levi-Civita tensor to the
Einstein-Hilbert action, coupled to a dynamical scalar field. It can
be thought of as an {\emph{effective theory}} (a low-energy
approximation to some more fundamental theory) whose signature is the
modification of parity asymmetric gravitational solutions, such as the
Kerr spacetime. It naturally arises as the low-energy approximation of
many different fundamental theories. The latter could be string theory
(i.e.,~Type I, IIb, Heterotic, etc.), where the inclusion of
the CS term is inescapable in the perturbative sector by the
Green-Schwarz anomaly-cancelling
mechanism~\cite{Polchinski:1998rr,Gates:2009pt}. In non-perturbative
string theory, CS corrections also arise in the presence of
D-instanton charges due to duality
symmetries~\cite{Alexander:2004xd}. The more fundamental theory could
also be loop quantum gravity, where the CS correction has also been
seen to generically arise when one of the fundamental constants of
that theory is promoted to a
field~\cite{Taveras:2008yf,Calcagni:2009xz,Mercuri:2009zt,Gates:2009pt}. Even
without recurring to a specific fundamental theory, the CS term
unavoidably arises in effective field theories as one of the only
non-vanishing, second-order curvature corrections to the
Einstein-Hilbert action in inflation~\cite{Weinberg:2008hq}.

The non-dynamical version of CS-modified gravity (where the scalar
field is non-evolving, but prescribed {\emph{a priori}}) has been
extensively studied and greatly constrained in the
past~\cite{Jackiw:2003pm,Alexander:2004wk,Alexander:2004xd,Alexander:2006mt,Guarrera:2007tu,Alexander:2007zg,Alexander:2007vt,Konno:2007ze,Alexander:2007kv,Alexander:2007qe,Yunes:2007ss,Smith:2007jm,Grumiller:2007rv,Grumiller:2008ie,Yunes:2008bu,Cantcheff:2008qn,Alexander:2008wi}. Only
recently, however, has there been a dedicated effort to analyze the
dynamical
theory~\cite{Yunes:2009hc,Konno:2009kg,Sopuerta:2009iy,Yunes:2009ry}.
In Ref.~\cite{Yunes:2009hc}, a small-coupling approximation was
developed to find the exterior gravitational field of a slowly
rotating compact object.  Perhaps rather
surprisingly~\cite{Psaltis:2007cw,Psaltis:2008bb}, the solution was
found to deviate from the Kerr spacetime because such a spacetime
sources a non-vanishing CS correction that couples to the spin-angular
momentum of the black hole. Later, in~\cite{Sopuerta:2009iy}, the Strong
Equivalence Principle was found to be satisfied in the dynamical theory and point particles
were indeed found to follow geodesics of the background spacetime.

In this paper, we calculate the CS-modified gravitational field inside
neutron stars and relate this to possible observables that would allow
us to constrain the theory. This analysis should be seen as
complimentary to that of~\cite{Yunes:2009hc}, thus completing the
gravitational field prescription everywhere in the spacetime of rotating
bodies. Our analysis is also similar to that of~\cite{Smith:2007jm},
except that here we focus on the dynamical theory instead of the
non-dynamical one and we employ a more realistic, tabulated
EOS. We shall see that if we employed the homogeneous
stellar structure equation used in~\cite{Smith:2007jm}, the
interior solution would not be CS-modified. Although in principle
there will be CS modifications to the Dirac equation that describes
the motion of fermions inside the star~\cite{Alexander:2008wi}, we
shall not consider those here.

We find that the main CS correction affects the gravitomagnetic sector
of the metric to leading order, as was found for the exterior solution~\cite{Yunes:2009hc}. 
Such a correction introduces modifications to the moment of inertia of the star, 
but not to the mass-radius relation to leading order. 
This implies that the usual degeneracy between alternative theory modifications and the EOS 
can be broken: measurements of the mass-radius relation from sub-kilohertz neutron stars can
be used to extract the EOS, while measurements of the
moment of inertia can be used to test the theory, given the EOS. 

Such a degeneracy break is sensitive to the spin-frequency of the neutron star observed. 
If the neutron star is spinning faster than $\Omega \sim 1 \; {\rm{kHz}}$, then second-order 
corrections in the spin-frequency to the mass-radius relation are not necessarily negligible 
and the degeneracy is not broken. One might then wonder whether such rapidly rotating neutron stars
exist in Nature. Up to date, most of the neutron stars observed spin slower than this limit 
(for a recent modeling of millisecond pulsars see~\cite{Story:2007xy}).
However, over the past year there has been a breakthrough in millisecond pulsar searches
through combined observations with the latest Fermi data. Such a breakthrough relies on
millisecond pulsars generically being strong $\gamma$-ray emitters~\cite{2009Sci...325..848A}. 
If so, this pulsar population could be calibrated through the diffuse 
$\gamma$-ray background they produce~\cite{2010JCAP...01..005F} 
(for an outline of such a new survey based on unidentified Fermi sources see~\cite{2010AAS...21545312R}). 
The tests we describe in this paper need not necessarily apply to this population, as then the degeneracy 
between mass-radius relation and CS modifications could not be easily broken.

Assuming that the degeneracy can be broken through observations of neutron stars with sub-kHz rotational
frequencies, we find that a $10\%$ accurate, moment of inertia measurement, as is expected in the near future 
with observations of the double pulsar~\cite{2005ApJ...629..979L,Kramer:2009zza}, could
place a constraint on the CS coupling constant of $\xi^{1/4} \lesssim
5 \; {\rm{km}}$. Such a constraint would be three-orders of magnitude stronger than the 
previous strongest bound~\cite{Yunes:2009hc}.

Before proceeding with the remainder of the paper, we first comment on
the relationship between neutron-star observations or tests and gravitational wave
ones. In the past, these two programs could have been seen as competing approaches, 
when in reality they should be seen as complementary. Gravitational wave observations
sample the fully dynamical, propagating sector of a theory, while neutron star observations 
test the strong-field, but stationary sector. Moreover, simultaneous gravitational wave-neutron star
observations, e.g.~of a merging neutron star binary, could yield the most stringent bounds
on alternative theories, as the combination of observations would help break degeneracies. 

This paper is divided as follows:
Sec.~\ref{ABC} defines the basics of CS modified gravity;
Sec.~\ref{NS-Tab-ds2-param} describes the metric parameterization and
the stress-energy description of the neutron star; 
Sec.~\ref{NS-GR} discusses CS modifications to neutron star models;
Sec.~\ref{Num-Sols} solves the CS modified equations of neutron star structure
numerically and present some results;
Sec.~\ref{Obs} computes some neutron star observables that are CS-modified
and discusses possible future constraints on the theory;
Sec.~\ref{Conclusions} concludes and points to future research.

We use the following conventions in this paper: we work exclusively in
four spacetime dimensions with signature
$(-,+,+,+)$~\cite{Misner:1973cw}; Latin letters $(a,b,\ldots,h)$ range
over all spacetime indices; round and square brackets around indices
denote symmetrization and anti-symmetrization respectively, namely
$T_{(ab)}=\frac12 (T_{ab}+T_{ba})$ and $T_{[ab]}=\frac12
(T_{ab}-T_{ba})$; partial derivatives are sometimes denoted by commas
(e.g.,~$\partial \theta/\partial r=\partial_r\theta=\theta_{,r}$),
while radial derivatives are sometimes denoted with an overhead prime
$\partial_{r} \omega = \omega'$ if $\omega = \omega(r)$; the notation
$A_{(m,n)}$ stands for a term of ${\cal{O}}(m,n)$, which itself stands
for terms of ${\cal{O}}(\epsilon^{m})$ or ${\cal{O}}(\zeta^{n})$; the
Einstein summation convention is employed unless otherwise specified;
finally, we use geometrized units where $G=c=1$ and where 
$M_{\odot} = 1.476 \; {{\rm km}}$ stands for a solar mass.

\section{The ABC of CS modified gravity}
\label{ABC}

In this section, we review some of the basics of CS modified gravity
(see the recent review article~\cite{Alexander:2009tp} for
further details).

\subsection{Basic Equations}

The CS action is defined as
\be
S \equiv S_{\rm EH} + S_{\rm CS} +  S_{\vartheta} + S_{\rm mat},
\ee
where 
\ba
\label{EH-action}
S_{\rm{EH}} &\equiv& \kappa \int_{{\cal{V}}} d^4x  \sqrt{-g}  R, 
\\
\label{CS-action}
S_{\rm{CS}} &\equiv& \frac{\alpha}{4} \int_{{\cal{V}}} d^4x  \sqrt{-g} \; 
\vartheta \; \pont\,,
\\
\label{Theta-action}
S_{\vartheta} &\equiv& - \frac{1}{2} \int_{{\cal{V}}} d^{4}x \sqrt{-g} \left[ g^{a b}
\left(\nabla_{a} \vartheta\right) \left(\nabla_{b} \vartheta\right) + 2 V(\vartheta) \right], \quad
\\
\label{mat}
S_{\textrm{mat}} &\equiv& \int_{{\cal{V}}} d^{4}x \sqrt{-g} {\cal{L}}_{\textrm{mat}}.
\ea
Equation~\eqref{EH-action} is the standard Einstein-Hilbert term;
Equation~\eqref{CS-action} is the CS correction;
Equation~\eqref{Theta-action} is the scalar-field term;
Equation~\eqref{mat} describes additional matter sources, with
${\cal{L}}_{\textrm{mat}}$ the matter Lagrangian density. The
quantity $\kappa^{-1} \equiv 16 \pi G$ is the coupling constant of GR,
while $\alpha$ is the CS coupling constant. This formulation assumes
that the CS scalar field $\vartheta$ is dimensionless, which then
forces $\alpha$ to have units of length squared\footnote{This
  formulation is equivalent to that of~\cite{Yunes:2009hc} with the
  choice $\beta \to 1$.}. As usual, $g$ is the determinant of the
metric, $\nabla_{a}$ is the covariant derivative associated with the
metric tensor $g_{ab}$, $R$ is the Ricci scalar, $\pont$ is the
Pontryagin density
\be
\label{pontryagindef}
\pont= R \tilde R = {\,^\ast\!}R^a{}_b{}^{cd} R^b{}_{acd}\,,
\ee
and the dual Riemann-tensor is  
\be
\label{Rdual}
{^\ast}R^a{}_b{}^{cd}\equiv \frac12 \epsilon^{cdef}R^a{}_{bef}\,,
\ee
with $\epsilon^{cdef}$ the 4-dimensional Levi-Civita tensor with sign
convention $\epsilon^{0123} = +1/\sqrt{-g}$, where $g$ is the determinant
of the metric.

The CS field equations are derived from the variation of the action
with respect to the metric tensor and the CS coupling field:
\ba
\label{eom}
R_{ab} + \frac{\alpha}{\kappa} C_{ab} &=& \frac{1}{2 \kappa} \left(T_{ab} - \frac{1}{2} g_{ab} T \right),
\\
\label{eq:constraint}
\square \vartheta &=& \frac{dV}{d\vartheta} - \frac{\alpha}{4} \pont,
\ea
with $R_{ab}$ the Ricci tensor, $\square \equiv \nabla_{a} \nabla^{a}$ the
D'Alembertian operator, and $C_{ab}$ the C-tensor
\be
\label{Ctensor}
C^{ab} \equiv \left(\nabla_c\vartheta\right) \epsilon^{cde(a}\nabla_eR^{b)}{}_d
+ \left(\nabla_c \nabla_{d}\vartheta\right) {\,^\ast\!}R^{d(ab)c}\,.
\ee
The total stress-energy tensor is
\be
\label{Tab-total}
T_{ab} \equiv T^{\textrm{mat}}_{ab} + T_{ab}^{\vartheta}, 
\ee
where $T^{\textrm{mat}}_{ab}$ stands for matter contributions,
and $T_{ab}^{\vartheta}$ is the scalar field contribution
\be
\label{Tab-theta}
T_{ab}^{\vartheta} 
\equiv \left(\nabla_{a} \vartheta\right) \left(\nabla_{b} \vartheta\right) 
    - \frac{1}{2}  g_{a b}\left(\nabla_{a} \vartheta\right) \left(\nabla^{a} \vartheta\right) 
-  g_{ab}  V(\vartheta).
\ee
We set the potential $V(\vartheta) = 0$ at the scales of interest,
with deviations possibly at the scale of supersymmetry breaking (see
e.g.,~the arguments presented in~\cite{Alexander:2009tp}).

Taking the divergence of Eq.~\eqref{eom}, the first term on the
left-hand side vanishes by the Bianchi identities, while the second
terms on both sides of this equation cancel each other because
\be
\label{nablaC}
\nabla_a C^{ab} = - \frac{1}{8} \left(\nabla^b \vartheta \right) \pont = \frac{1}{2 \kappa} \nabla^{a} T_{ab}^{\vartheta},
\ee
provided that Eq.~\eqref{eq:constraint} is satisfied. This then
establishes that $\nabla_{a} T^{ab}_{\textrm{mat}} = 0$ in dynamical
CS modified gravity, thus forcing point particles to move on
geodesics.

Two versions of CS modified gravity exist: the non-dynamical version
and the fully dynamical one. In the former, the scalar field is an {\emph{a
priori}}, prescribed function that lacks dynamics. With our choice of
coupling normalization, this theory can be reproduced in the limit
$\alpha \to \infty$, as then the scalar field evolution equation
becomes $\pont = 0$ (the so-called Pontryagin constraint), which
restricts the space of allowed solutions. We shall not consider this
model here, as it has been heavily constrained by solar system and
binary pulsar
observations~\cite{Alexander:2007vt,Alexander:2007zg,Smith:2007jm,Alexander:2007:gwp,Yunes:2008bu,Yunes:2008ua,Alexander:2009tp}. Instead,
we shall concentrate on the fully dynamical theory, whose equations 
of motion are given in Eqs.~\eqref{eom} and \eqref{eq:constraint}.

\subsection{Perturbative Constraints in \\ Dynamical CS Modified Gravity}
\label{pert-const}

The main idea of perturbative constraints (see
e.g.,~\cite{Jaen:1986iz,Eliezer:1989cr,Simon:1990ic,DeDeo:2007yn,Cooney:2008wk,Cooney:2009rr})
is to perform a small-coupling expansion of the solution to a certain
set of differential equations. From a mathematical standpoint, this is nothing
but an application of asymptotic analysis or perturbation theory (see e.g.,~\cite{BO}).

In the context of CS modified gravity,
perturbative constraints reduce to the {\emph{small-coupling}}
approximation~\cite{Yunes:2009hc}, where one searches for solutions of Eqs.~\eqref{eom}
and~\eqref{eq:constraint} in the limit $\zeta \ll 1$, where $\zeta$ is
of ${\cal{O}}[\alpha^{2}/({\cal{L}}^{4} \kappa)]$ and ${\cal{L}}$ is
some characteristic scale that describes the physical system under
consideration. For example, when searching for spinning black hole
solutions~\cite{Yunes:2009hc}, ${\cal{L}}_{\rm BH} = M_{\rm BH}$ 
with $M_{\rm BH}$ the black hole mass and $\zeta = \alpha^{2}/(\kappa M^{4})$. 
In the case of neutron stars, the natural length scale is the neutron star radius 
${\cal{L}}_{\rm NS}  = R$ and $\zeta = \alpha^{2}/(\kappa R^{4})$. 

In addition to the use of the perturbative constraint scheme (i.e.,~the small-coupling
approximation), we additionally focus on {\emph{slowly-rotating}} solutions. We
define slow-rotation by requiring that $\epsilon \ll 1$, where $\epsilon =  {\cal{O}}({\cal{L}} \; {\it{\Omega}})$
and $\it\Omega$ is the angular velocity of the physical system. As before,
since for neutron stars ${\cal{L}}_{\rm NS} = R$, then $\epsilon = {\cal{O}}(R \; \Omega)$, 
where $\Omega$ is the star's angular velocity.

The quantities $\epsilon$ and $\zeta$ are {\emph{book-keeping}} perturbative parameters
associated with the slow-rotation and small-coupling approximations, respectively. 
We shall use them to remind ourselves of the order of the approximation. Note
that $\zeta \sim \alpha^{2}$ instead of $\sim \alpha$ because the
evolution equation for $\vartheta$ forces this field to be linear in
$\alpha$, and thus, the C-tensor becomes linear in $\alpha$ as well,
leading to corrections to the field equations that scale as
$\alpha^{2}$.

Combining the small-coupling and slow-rotation approximations leads to {\emph{bivariate}} or
{\emph{two-parameter}} expansions. Schematically, all fields $A$ are
expanded as
\be
A = \sum_{m,n} \epsilon^{m} \zeta^{n} A_{(m,n)},
\ee
where $A_{(m,n)}$ is a term of ${\cal{O}}(\epsilon^{m},\zeta^{n})$, which means 
a term of ${\cal{O}}(\epsilon^{m})$ and ${\cal{O}}(\zeta^{n})$. For example, a term 
$A_{(1,1)}$ is proportional to $\epsilon \; \zeta$ and thus of ${\cal{O}}(\epsilon,\zeta)$,
while a term $A_{(2,1)}$ is proportional to $\epsilon^{2} \; \zeta$ and thus of ${\cal{O}}(\epsilon^{2},\zeta)$
[clearly $A_{(2,1)}$ is negligible relative to $A_{(1,1)}$].  
The metric tensor can be expanded bivariately as
\ba
g_{ab} &=& g_{ab}^{(0,0)} + \zeta g^{(0,1)}_{ab} + \epsilon g^{(1,0)}_{ab} 
\nonumber \\
&+& \zeta^{2} g^{(0,2)}_{ab} + \epsilon^{2} g^{(2,0)}_{ab} + \epsilon \zeta g^{(1,1)}_{ab}  + {\cal{O}}(3,3),
\label{small-cou-metric}
\ea
where $g_{ab}^{(0,0)}$ is some background metric that satisfies the
Einstein equations, while $(g_{ab}^{(0,1)},g_{ab}^{(1,0)})$ are first
order perturbations and
$(g_{ab}^{(0,2)},g_{ab}^{(2,0)},g_{ab}^{(1,1)})$ are second-order
perturbations. Given an exact solution to the Einstein equations,
$g_{ab}^{\rm GR}$, its expansion in $\epsilon$ must satisfy
\be
g_{ab}^{\rm GR} = g_{ab}^{(0,0)} + \epsilon g_{ab}^{(1,0)} + \epsilon^{2} g_{ab}^{(2,0)} + {\cal{O}}(\epsilon^{3}).
\ee
Thus, all terms $g_{ab}^{(N,0)}$ for all $N$ are pieces of the
slow-rotation expansion of a GR solution, while terms that are
proportional to $\zeta$ represent CS corrections.

The scalar field can also be expanded bivariately via
\ba
\vartheta_{ab} &=& \vartheta_{ab}^{(0,0)} + \zeta \vartheta^{(0,1)}_{ab} + \epsilon \vartheta^{(1,0)}_{ab} 
\nonumber \\
&+& \zeta^{2} \vartheta^{(0,2)}_{ab} + \epsilon^{2} \vartheta^{(2,0)}_{ab} + \epsilon \zeta \vartheta^{(1,1)}_{ab}  + {\cal{O}}(3,3).
\label{small-cou-field}
\ea
Each piece in this expansion is determined by the evolution equation
for the scalar field [Eq.~\eqref{eq:constraint}], which in turn
depends on the Pontryagin
density. In~\cite{Grumiller:2007rv,Yunes:2009hc}, it was shown that for
spherically symmetric spacetimes $\pont = 0$, thus forcing any
physical CS scalar field to a constant and allowing us to set
$\vartheta^{(0,N)} = 0$ for all $N$.  Angular momentum breaks the
spherical symmetry and forces $\pont \sim \Omega \sim \epsilon$ to leading order,
which then allows us to write
\ba
\vartheta_{ab} &=& \epsilon \vartheta^{(1,0)}_{ab}  + \epsilon^{2} \vartheta^{(2,0)}_{ab} + \epsilon \zeta \vartheta^{(1,1)}_{ab}  + {\cal{O}}(3,3).
\ea

We see that, to leading order, the C-tensor is linear in $\epsilon$,
which then forces the leading order correction to the metric to be
proportional to $(\epsilon \; \zeta)$, i.e.,~a second-order correction in
perturbation theory. The metric expansion simplifies to
\be
g_{ab} = g_{ab}^{\rm GR} + \epsilon \; \zeta \; g_{ab}^{(1,1)} + {\cal{O}}(3,3).
\ee
Solving the modified field equations reduces simply to solving for $g_{ab}^{(1,1)}$. 

The combination of the small-coupling and slow-rotation approximations
establishes a well-defined iteration scheme. First, we solve the
Einstein equations to linear order in $\epsilon$ to obtain
$g^{(0,0)}_{ab} + \epsilon g^{(1,0)}_{ab}$. Second, we use this metric
to compute $\pont_{(1,0)}$ and from this solve
Eq.~\eqref{eq:constraint} for $\vartheta^{(1,0)}$.  Third, we use this
scalar field and the $g_{ab}^{(0,0)}$ piece of the metric to compute
the C-tensor, which when combined with the Einstein tensor
$G_{ab}^{(1,1)}$ leads to differential equations for $g^{(1,1)}_{ab}$.
Fourth, we solve these differential equations and iterate the scheme
to higher order (see, e.g.,~Ref.~\cite{Yunes:2009hc} for more details
on this boot-strapping technique). In this paper, we work to leading order
only.

\section{Metric Parameterization and the Stress-Energy Tensor of Neutron-Star Matter}
\label{NS-Tab-ds2-param}

In this section, we consider the parameterization of the metric tensor and
the description of the stress-energy tensor for neutron-star matter.
We shall search for solutions to the field equations for a metric of
the form~\cite{1967ApJ...150.1005H}
\be
ds^{2} = - e^{2 \lambda} dt^{2} + e^{2 \beta} dr^{2} + r^{2} e^{2 \gamma}\left[d\theta^{2} + e^{2 \chi} \sin^{2}{\theta} \left(d\phi - \omega dt\right)^{2} \right],
\label{gen-lin-el}
\ee
where $(r,\theta,\phi)$ are (Boyer-Lindquist) spherical polar coordinates and
$(\lambda,\beta,\gamma,\chi,\omega)$ are independent functions of radius
and polar angle $(r,\theta)$.  The function $\omega$ is effectively
the angular velocity of an observer falling in from infinity and, as
such, is at least linearly proportional to the angular velocity of
the rotating star.

We decompose the metric functions in bivariate expansions as 
\ba
\lambda &=& \lambda_{(0,0)}(r) + \epsilon^{2} \; \lambda_{(2,0)}(r,\theta) + \epsilon \zeta \; \lambda_{(1,1)}(r,\theta), 
\nonumber \\
\beta &=& \beta_{(0,0)}(r) + \epsilon^{2} \; \beta_{(2,0)}(r,\theta) + \epsilon \zeta \; \beta_{(1,1)}(r,\theta),
\nonumber \\
\gamma &=& \gamma_{(0,0)}(r) + \epsilon^{2} \; \gamma_{(2,0)}(r,\theta) + \epsilon \zeta \; \gamma_{(1,1)}(r,\theta), 
\nonumber \\
\chi &=& \chi_{(0,0)}(r) + \epsilon^{2} \; \chi_{(2,0)}(r,\theta) + \epsilon \zeta \; \chi_{(1,1)}(r,\theta), 
\nonumber \\
\omega &=& \epsilon \; \omega_{(1,0)}(r) + \epsilon \zeta \; \omega_{(1,1)}(r,\theta), 
\label{omega-exp}
\ea
To linear order in $\epsilon$, we can safely assume that all
functions of ${\cal{O}}(0,0)$ and ${\cal{O}}(1,0)$ depend only on
radius. Notice that we have set
$(\lambda_{(0,1)},\beta_{(0,1)},\gamma_{(0,1)},\chi_{(0,1)})=0$ and
$(\lambda_{(0,2)},\beta_{(0,2)},\gamma_{(0,2)},\chi_{(0,2)})=0$ as the
C-tensor corrections start at ${\cal{O}}(1,1)$, since there cannot be
CS-deformations that are angular momentum independent due to parity
invariance. Notice further that we have also set
$(\omega_{(0,0)},\omega_{(2,0)})=0$ and
$(\lambda_{(1,0)},\beta_{(1,0)},\gamma_{(1,0)},\chi_{(1,0)})=0$
because the GR gravitomagnetic and diagonal sectors are known to be
series in odd and even powers of the angular velocity respectively in
the slow-rotation approximation~\cite{1967ApJ...150.1005H}.  Finally,
we set $\gamma_{(0,0)} = 1 = \chi_{(0,0)}$, which amounts to a
redefinition of the radial coordinate.

In order to get a feel for how these functions behave, we first
present the exterior gravitational field of a neutron star in GR.
This exterior field is given by the slow-rotation expansion of the
Kerr metric, where the small-rotation parameter $\epsilon$ is now of
${\cal{O}}(M \; a)$, with $M$ the mass of the compact object and $a$
its Kerr parameter. Notice that the Kerr parameter is related to the
angular momentum of the compact object via $|J| = M a$, which then 
implies $a = I \; \Omega/M$, since by definition $J = I \; \Omega$, where
$I$ is the moment of inertia (to be defined properly later).  To zeroth order
in the spin, we find
\ba
\left(e^{2 \lambda}\right)_{(0,0)} &=& 1 - \frac{2 M}{r},
\qquad
\left(e^{2 \beta}\right)_{(0,0)} = \left(1 - \frac{2 M}{r} \right)^{-1},
\nonumber \\
\left(e^{2 \gamma}\right)_{(0,0)} &=& 1,
\qquad 
\left(e^{2 \chi}\right)_{(0,0)} = 1,
\qquad
\omega_{(0,0)} = 0.
\label{BC-00}
\ea
To linear order we find
\ba
\left(e^{2 \lambda}\right)_{(1,0)} &=& 0 = \left(e^{2 \beta}\right)_{(1,0)},
\nonumber \\
\left(e^{2 \gamma}\right)_{(1,0)} &=& 0 = \left(e^{2 \chi}\right)_{(1,0)}, 
\qquad 
\omega_{(1,0)} = \frac{2 M a}{r^{3}}.
\label{BC-10}
\ea
To second order we find
\ba
\left(e^{2 \lambda}\right)_{(2,0)} &=&  \frac{4 M^{2} a^{2}}{r^{4}} \sin^{4}{\theta} + \frac{2 M a^{2}}{r^{3}} \cos^{2}{\theta},
\nonumber \\
\left(e^{2 \beta}\right)_{(2,0)} &=& \frac{a^{2}}{r^{2}} \left(1 - \frac{2 M}{r} \right)^{-1} 
\left[\cos^{2}{\theta} - \left(1 - \frac{2 M}{r} \right)^{-1} \right],
\nonumber \\
\left(e^{2 \gamma}\right)_{(2,0)} &=& \frac{a^{2}}{r^{2}} \cos^{2}{\theta},
\nonumber \\
\left(e^{2 \chi}\right)_{(2,0)} &=&  \frac{a^{2}}{r^{2}} \left(1 + \frac{2 M}{r} \right) \sin^{2}{\theta},
\qquad
\omega_{(2,0)} = 0,
\label{BC-20}
\ea

We model the stress-energy tensor of matter in the neutron star 
as that of a perfect fluid
\be
T_{ab}^{\rm mat} = \left( \rho + p \right) u_{a} u_{b} + p g_{ab},
\label{pf}
\ee
where $p \equiv p(r)$ and $\rho \equiv \rho(r)$ are the pressure and
density, while $u^{a} = \left(u^{0},0,0,\Omega u^{0}\right)$ is the
fluid's four-velocity that corresponds to a constant angular velocity
$\Omega$ (we do not consider differentially rotating stars here).  
Normalization of this four-velocity requires $u_{a} u^{a} = -1$, and hence
\ba
u^{0} &=& e^{- \lambda_{(0,0)}} \left[1 - \epsilon^{2} \lambda_{(2,0)} - \epsilon \zeta \; \lambda_{(1,1)} 
\right. 
\nonumber \\
&+& \left.
\frac{1}{2} \epsilon^{2} e^{-2 \lambda_{(0,0)}} r^{2} \sin^{2}{\theta} \left(\Omega - \omega_{(1,0)} \right)^{2} \right].
\ea
We now expand the stress-energy tensor order by order in $\epsilon = R
\Omega \ll 1$. To zeroth-order, the only non-vanishing components are
\ba
{}^{(0,0)}T_{tt}^{\rm mat} &=& \rho e^{2 \lambda^{(0,0)}},
\qquad
{}^{(0,0)}T_{rr}^{\rm mat} = p \; e^{2 \beta^{(0,0)}},
 \\ \nonumber
{}^{(0,0)}T_{\theta \theta}^{\rm mat} &=& p r^{2} = \frac{{}^{(0,0)}T_{\phi\phi}^{\rm mat}}{\sin^{2}{\theta}}.
\ea
To first order in $\epsilon$, the non-vanishing terms are
\be
{}^{(1,0)}T_{t \phi}^{\rm mat} = - r^{2} \sin^{2}{\theta} \left[\rho \left(\Omega - \omega^{(1,0)} \right) + p \Omega \right],
\ee
while to second order in $\epsilon$ we find
\ba
{}^{(2,0)}T_{tt}^{\rm mat} &=& 2 \; \lambda_{(2,0)} \; \rho \; e^{2 \lambda^{(0,0)}} +  r^{2} \sin^{2}{\theta} \left[\rho \left(\Omega^{2} - \omega^{2}_{(1,0)} \right) 
\right.
\nonumber \\
&+& \left. p \Omega^{2} \right]
+ 2 \; p\; r^2 \sin^{2}{\theta} \left(\gamma_{(2,0)} + \chi_{(2,0)} \right),
\nonumber \\
{}^{(2,0)}T_{rr}^{\rm mat} &=& 2 \; p \; e^{2 \beta^{(0,0)}} \beta_{(2,0)},
\nonumber \\
{}^{(2,0)}T_{\phi\phi}^{\rm mat} &=& e^{-2\lambda_{(0,0)}} \; r^4 \left(\omega^{(1,0)}-\Omega\right)^2 \left(p+\rho\right) \sin^{4}{\theta}.
\ea
The cross terms of ${\cal{O}}(1,1)$ are
\ba
{}^{(1,1)}T_{tt}^{\rm mat} &=& 2 \; \lambda_{(1,1)} \; \rho \; e^{2 \lambda^{(0,0)}},
\nonumber \\
{}^{(1,1)}T_{rr}^{\rm mat} &=& 2 \; p \; e^{2 \beta^{(0,0)}} \beta_{(1,1)},
\nonumber \\
{}^{(1,1)}T_{t\phi}^{\rm mat} &=& r^2 \rho \; \omega_{(1,1)} \sin^{2}{\theta}.
\nonumber \\
{}^{(1,1)}T_{\phi\phi}^{\rm mat} &=& 2 \; p\; r^2 \sin^{2}{\theta} \left(\gamma_{(1,1)} + \chi_{(1,1)} \right).
\ea
The trace is always $T^{\rm mat} \equiv g^{a b} T_{ab}^{\rm mat} = 3 p - \rho$ up to
${\cal{O}}(\epsilon^{3})$. Notice that the density and the pressure
are assumed to be independent of $\Omega$ and $\alpha$, and thus they
carry no $\epsilon$ or $\zeta$ dependence.

The components of the stress-energy obtained in this section depend only on the perfect-fluid
assumption of Eq.~\eqref{pf} and the stationarity and axisymmetry assumption of the metric 
of Eq.~\eqref{gen-lin-el}. In general, for an arbitrary metric tensor, all components of the stress-energy
will be non-vanishing, but here many components vanish due to the axisymmetry condition. Moreover,
in the presence of electromagnetic fields, there will be additional contributions that are not accounted for
here.

\begin{figure}[htb]
\includegraphics[width=8cm,clip=true]{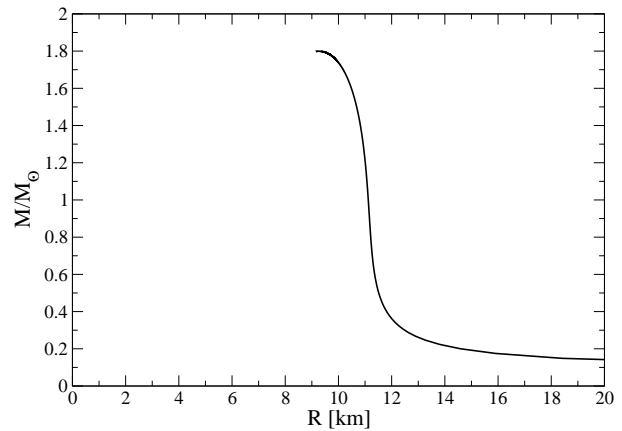} 
\caption{\label{fig.eos} 
The mass-radius relation for neutron stars calculated using the FPS
EOS. Notice that for the FPS EOS, neutron stars are smaller than $2 M_{\odot}$
and have radii $\sim 10$ km.}
\end{figure}

The EOS determines the relation between the density and
the pressure: $p = p(\rho)$. We shall employ here the FPS 
EOS~\cite{Friedman:1981qw}, which predicts neutron-star masses
smaller than $2 M_\odot$ and radii ($\simeq 10$~km; see
Fig.~\ref{fig.eos}) that are consistent with recent observations of
X-ray bursters. 

\section{Neutron Stars in Dynamical CS Modified Gravity}
\label{NS-GR}

In this section, we solve the modified field equations in the
small-coupling and slow-rotation approximation (within the
perturbative constraint framework).  Each subsection considers the
equations at each successive order in the perturbation theory ladder.

\subsection{Zeroth-Order Field Equations}

We first study the field equations to ${\cal{O}}(0,0)$, i.e.,~the
solutions that are independent of $\zeta$ and $\epsilon$.  To this
order, the C-tensor can be neglected, as it is at least linear in
$\zeta$ and $\epsilon$. The field equations then reduce exactly to
Einstein's equations (for a review of neutron star solution in GR see
e.g.~\cite{Lattimer:2000nx,1996csnp.book.....G,2001LNP...578.....B}).
The Einstein tensor possesses only three independent components:
$(t,t)$, $(r,r)$ and $(\theta,\theta)$. Using the definition $m(r)
\equiv r (1 - e^{-2 \beta})/2$, the $(t,t)$ and the $(r,r)$ components reduce to the mass conservation
equation and an equation for the $\lambda$ metric function,
respectively
\ba
\label{Mprime-eq}
\frac{2 m^\prime_{(0,0)}}{r^{2}} &=& 8 \pi \rho,
\\
\lambda^\prime_{(0,0)} &=& \frac{ m_{(0,0)} + 4 \pi p r^3}{r (r - 2 m_{(0,0)})}.
\label{alphaprime-eq}
\ea
Hereafter, a prime stands for differentiation with respect to radius. We have explicitly included
the index $(0,0)$ here to remind us that we are dealing with the
${\cal O}(0,0)$ coefficients of the metric expansion.

The system of differential equations becomes complete once we specify
a third differential equation relating the rate of change of pressure to the density. We
could use the $(\theta,\theta)$ component, but the equations simplify
the most if we instead use the equations of stress-energy conservation
(which hold to all orders in dynamical CS modified gravity):
$\nabla^{\mu} T_{\mu \nu}^{\rm mat} = 0$. The $r$-component of this equation
becomes
\be
\nabla^{\mu} T_{\mu r}^{\rm mat} = \lambda'_{(0,0)} \left(\rho + p\right) + p' = 0.
\ee
Combining this equation with Eq.~\eqref{alphaprime-eq} we obtain the
Tolman-Oppenheimer-Volkoff (TOV) equation
\be
\label{TOV-eq}
p' = -  \frac{ m_{(0,0)} + 4 \pi p r^3}{r (r - 2 m_{(0,0)})} 
   \left( \rho + p \right).
\ee
Given an EOS relating density and pressure, we can then solve the
system of equations~\eqref{Mprime-eq}-\eqref{TOV-eq}.  

\subsection{Zeroth-Order Scalar-Field Evolution Equation}
\label{0th-scal-evol}

Once we have obtained the metric functions $\lambda_{(0,0)}$ and
$\beta_{(0,0)}$, we solve the evolution equation for the CS scalar
field to zeroth order in $\epsilon$ and $\zeta$. To this order,
however, the spacetime is completely spherically
symmetric. In~\cite{Grumiller:2007rv}, it was shown that for such
spacetimes, the Pontryagin density identically vanishes, i.e.,~$\pont
= 0$, which then forces the evolution equation to become
\be
\square \vartheta^{(0,0)} = 0\;.
\ee
Here again the $(0,0)$ index reminds us that we are working to
zeroth-order in both perturbation parameters.

Since we are searching for a stationary and axisymmetric solution, 
$\vartheta$ can depend only on $(r,\theta)$.
With this assumption, the above differential equation becomes
\ba
\square \vartheta_{(0,0)} &=& 
e^{-2 \beta^{(0,0)}} \vartheta_{,rr}^{(0,0)}
+ \vartheta_{,r}^{(0,0)} e^{-2 \beta^{(0,0)}} \left(\lambda'_{(0,0)} - \beta'_{(0,0)}
\right.
\nonumber \\
&+& \left. \frac{2}{r} \right) 
+ \frac{1}{r^{2}} \vartheta_{,\theta\theta}^{(0,0)} + \frac{\cot{\theta}}{r^{2}} \vartheta_{,\theta}^{(0,0)}.
\label{Dal}
\ea
In fact, the above differential equation also holds to
${\cal{O}}(1,0)$ with the substitution $\vartheta^{(0,0)} \to
\vartheta^{(1,0)}$.

This differential equation only possesses ill-behaved solutions,
i.e.,~solutions that lead to a scalar field with infinite energy.  Such
an observation has in fact already been made when investigating the
exterior gravitational field of slowly-rotating black
holes~\cite{Yunes:2009hc}. The only solution that is consistent with a
scalar field with finite energy is $\vartheta^{(0,0)} = \rm{const.}$,
which is the choice we make here. Of course, such a scalar field leads
to no modification to the field equations as the C-tensor depends on
$\vartheta$ derivatives.

\subsection{First-Order Field Equations}

We now search for solutions of the modified field equations that are
either linear in $\epsilon$ or linear in $\zeta$, i.e.,~solutions to
${\cal{O}}(1,0)$ or ${\cal{O}}(0,1)$. As already argued, however, there 
are no CS corrections independent of $\epsilon$,  and thus $\omega_{(0,1)} = 0$. 
As for corrections of ${\cal{O}}(1,0)$, the C-tensor does not contribute, while 
the Einstein tensor contains only one equation, the $(t,\phi)$ component, which reduces 
in normal form to
\be
\omega''_{(1,0)} + D \omega'_{(1,0)} + V \omega_{(1,0)} 
= - 16 \pi \Omega \frac{\rho + p}{1 - 2 m_{(0,0)}/r}.
\label{omega-eq}
\ee
Here, the dissipative and potential term respectively are
\ba
D &\equiv& \frac{4}{r} - \left(\lambda'_{(0,0)} + \beta'_{(0,0)}\right),
\\
V &\equiv& 2 \beta'_{(0,0)} \lambda'_{(0,0)} - \frac{2}{r} \left(\lambda'_{(0,0)} - \beta'_{(0,0)} \right)
\nonumber \\
&-& 2 \lambda'^{2}_{(0,0)} - 2 \lambda''_{(0,0)} - 16 \pi e^{2 \beta_{(0,0)}} \rho.
\ea
We can simplify the above expressions using the zeroth order ones in
Eqs.~\eqref{Mprime-eq}-\eqref{TOV-eq}.  Doing so we find
\ba
D &=& \frac{4}{r} - \frac{4 \pi r^2  \left( \rho + p\right)}{r - 2 m_{(0,0)}},
\\
V &=& -\frac{16 \pi r}{r - 2 m_{(0,0)}} \left(\rho+p\right).
\ea
We consider some analytic solutions to this equations in the Appendix.
Once more, once $\lambda$, $\beta$, $p$ and $\rho$ are known as 
functions of radius (from the previous rung of the perturbative ladder), the
above equations can be solved to find $\omega_{(1,0)}$ as a function of $r$. 

\subsection{First-Order Scalar Field Evolution Equation}

We now consider the evolution equation for the scalar field to
${\cal{O}}(1,0)$ or ${\cal{O}}(0,1)$, i.e.,~we search for an evolution equation for this
scalar field that is linear in $\epsilon$ or linear in $\zeta$. The
left-hand side of the evolution equation is given by Eq.~\eqref{Dal}, as we showed in
Sec.~\ref{0th-scal-evol}. The right-hand side of the evolution
equation depends on the Pontryagin density, which to linear order is:
\ba
\pont_{(1,0)} &=& \frac{8}{r^{2}} \cos{\theta} \; e^{-3 \beta_{(0,0)} - \lambda_{(0,0)}} 
\left[ 1 - e^{2 \beta_{(0,0)}} 
\right.  
\nonumber \\
&+& \left.
r \left(\beta'_{(0,0)} - \lambda'_{(0,0)}\right) + r^{2} \left(\lambda'^{2}_{(0,0)} - \lambda'_{(0,0)} \beta'_{(0,0)} 
\right. \right. 
\nonumber \\
&+& \left. \left.
\lambda''_{(0,0)} \right) \right] \; \omega'_{(1,0)}
\ea
We see immediately that the Pontryagin density is of ${\cal{O}}(1,0)$ and there is no ${\cal{O}}(0,1)$ piece, 
which immediately implies we can neglect $\vartheta_{(0,1)}$.

The complicated source for the evolution equation of the scalar field
is greatly simplified once we make use of the lower-order
equations. Inserting Eqs.~\eqref{Mprime-eq}-\eqref{TOV-eq}, the
Pontryagin density simplifies to
\ba
\pont_{(1,0)} &=& -\frac{48}{r^{3}} e^{-\lambda_{(0,0)}} \cos{\theta} 
\left(m_{(0,0)} -  \frac{4 \pi}{3} r^{3} \rho \right) 
\nonumber \\
&\times& \sqrt{1 - \frac{2 m_{(0,0)}}{r}} \; \omega'_{(1,0)} .
\label{pont-eq}
\ea
where $\omega'_{(1,0)}$ is to be understood as a known function
obtained after solving Eq.~\eqref{omega-eq}. For completeness, we
present the full evolution equation for the scalar field
[i.e.,~Eq.~\eqref{eq:constraint} to first order in $\epsilon$] below:
\ba
&& e^{-2 \beta_{(0,0)}} \vartheta_{,rr}^{(1,0)}
+ \vartheta_{,r}^{(1,0)} e^{-2 \beta_{(0,0)}} \left(\lambda'_{(0,0)} - \beta'_{(0,0)} + \frac{2}{r} \right) 
\nonumber \\
&+&  \frac{1}{r^{2}} \vartheta_{,\theta\theta}^{(1,0)} +\frac{\cot{\theta}}{r^{2}} \vartheta_{,\theta}^{(1,0)}
= \frac{12 \alpha}{r^{3}} e^{-\lambda_{(0,0)}} \cos{\theta} \left(m_{(0,0)}  
\right.
\nonumber \\
&& \left. \qquad \qquad \qquad \qquad 
-  \frac{4 \pi}{3} r^{3} \rho \right) 
\sqrt{1 - \frac{2 m_{(0,0)}}{r}} \; \omega'_{(1,0)}
\nonumber \\
\label{scal-evol-first-O}
\ea

Before proceeding with the solutions, it is worth studying the
structure of the Pontryagin constraint. The quantity in between
parenthesis in Eq.~\eqref{pont-eq} is simply
\be
m_{(0,0)} -  \frac{4 \pi}{3} r^{3} \rho = \frac{4 \pi}{3} r^{3} \left(\bar{\rho} - \rho\right),
\label{CS-term-Eq}
\ee
where we have defined the mean density $\bar{\rho} \equiv 3
m_{(0,0)}/(4 \pi r^{3})$. Clearly, for constant density stars, the
above correction identically vanishes and the leading order term of
the Pontryagin density is of ${\cal{O}}(\epsilon^{3})$. This is indeed
the case for the EOS employed in~\cite{Smith:2007jm}, which would 
incorrectly suggest that for such an EOS there is no observable CS 
correction to neutron stars. Note here that there is no $\epsilon^{2}$
contribution to $\pont$ because a term of this order is parity invariant,
while only odd-powers of $\epsilon$ can break this symmetry. 

\begin{figure}[t]
\epsfig{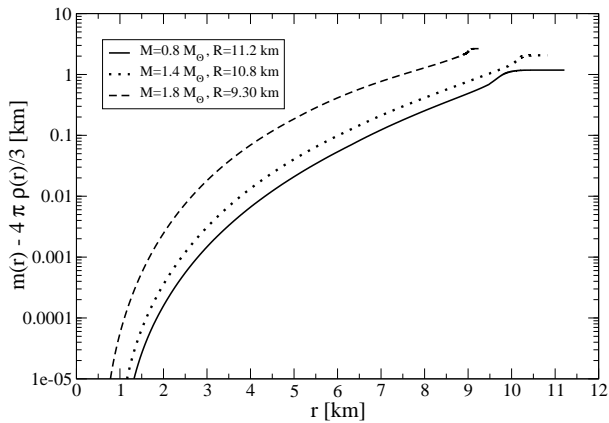}
\caption{\label{fig.1} The CS sourcing function $m_{(0,0)} - 4
  \pi r^{3} \rho/3$ as a function of radius in km.  Close to the core,
  the function vanishes, while it reaches its maximum near the
  surface. The three curves correspond to stars with different central
  densities and hence with different masses.}
\end{figure}

For non-constant density EOSs, one expects the above
density-dependent quantity to be close to zero near the neutron star core and reach
its maximum value close to its surface, a behavior that
is verified in Fig.~\ref{fig.1}. This figure shows the relevant term
as a function of radius in units of kilometers. The three curves
correspond to three different central densities that lead to different
total masses. Clearly, in the core (near zero radius), the Pontryagin
density is nearly vanishing as the density is nearly constant, while
it reaches its maximum near the surface of the star, which is close in
magnitude to the total mass of the star.  This behavior implies that
the scalar field will be driven to non-vanishing values near the
surface, which then in turn will lead to modifications to the
gravitational field near the surface of the star.  Due to this
density dependence, the CS correction is not very sensitive to
uncertainties in the nuclear physics, as the EOS near the neutron star
surface is fairly robust.

A physical reason behind the vanishing of $\pont$ for constant-density stars
is the following. In~\cite{Grumiller:2007rv}, it was shown that $\pont = E_{ab} B^{ab}$,
where $E_{ab}$ and $B_{ab}$ are the electric and magnetic tidal tensors, which
can be constructed from the Weyl tensor. Such a relation allows for an analogy between
CS modified gravity and electromagnetic theory. In the latter, we know that the equations
of motion for electromagnetic radiation satisfy wave equations in vacuum, but in the presence
of matter, these equations acquire sources, e.g.~$\square E \propto \nabla \rho + \dot{J}$
and $\square B \propto \nabla \times J$, where $\rho$ and $J$ are the charge and current densities
respectively. Thus, in the presence of matter, the inhomogeneous solutions to the electric and magnetic
fields need not be perpendicular to each other, leading to a non-vanishing $E \cdot B$, and, by analogy,
a non-vanishing Pontryagin density. It is no surprise then that $\pont$ is proportional to derivatives
of the matter density and the current, the latter of which in the gravitational sector is proportional to
derivatives of the $\omega$ metric perturbation.

\subsection{Second-Order Field Equations}
\label{2nd-O-fieldeqs}

The field equations to second order can be split into two sets: that
of ${\cal{O}}(1,1)$ and that of ${\cal{O}}(2,0)$ (as argued before, there
is no correction to the metric of ${\cal{O}}(0,2)$). The latter reduces
simply to the second-order GR equations of structure, which we do not
present here. This is a result of the C-tensor vanishing identically to
${\cal{O}}(2,0)$. The ${\cal{O}}(1,1)$ equations contain CS
modifications, as the C-tensor does not vanish in this
case.

The only surviving term of the C-tensor is
\ba
&& \left( \frac{\alpha}{\kappa} C_{t \phi}\right)^{(1,1)} =  - 4 \pi \sin{\theta}  \; e^{\lambda_{(0,0)}} \left(1 - \frac{2 m_{(0,0)}}{r} \right)^{1/2}
\\ \nonumber
&& \qquad \qquad \quad 
\left[ \left(\frac{\bar{\rho} - \rho}{r} + \frac{d \rho}{dr} \right) \partial_{\theta} \vartheta_{(1,0)} 
- \left(\bar{\rho}-\rho\right) \partial_{r \theta} \vartheta_{(1,0)} \right].
\ea
Since all quantities in this equation have already been computed, the
C-tensor effectively acts as a new source for the field equations (a new
effective matter term). The accompanying component of the Einstein tensor to this order is
given by
\ba G_{t \phi}^{(1,1)} &=& \frac{r^{2}}{2} \sin^{2}{\theta} \left(1 -
\frac{2 m_{(0,0)}}{r} \right) \; \partial_{rr} \omega_{(1,1)}
\nonumber \\ &+& \sin^{2}{\theta} \left[2 r \left(1 - \frac{2
    m_{(0,0)}}{r}\right) - 2 \pi r^{3} \left( \rho + p \right) \right]
\; \partial_{r} \omega_{(1,1)} \nonumber \\ &-& 8 \pi p r^{2}
\sin^{2}{\theta} \; \omega_{(1,1)} + \frac{3}{2} \sin{\theta}
\cos{\theta} \; \partial_{\theta} \omega_{(1,1)} \nonumber \\ &+&
\frac{1}{2} \sin^{2}{\theta} \; \partial_{\theta \theta}\omega_{(1,1)}\;.
\ea
The stress-energy tensor to this order has already been given in a
previous section. The full $(t,\phi)$ component of the modified field
equations to ${\cal{O}}(1,1)$ is then
\begin{widetext}
\ba
&& \frac{r^{2}}{2} \left(1 - \frac{2 m_{(0,0)}}{r} \right) \; \partial_{rr} \omega_{(1,1)}
+ \left[2 r \left(1 - \frac{2 m_{(0,0)}}{r}\right) - 2 \pi r^{3} \left( \rho + p \right) \right] \; \partial_{r} \omega_{(1,1)}
+ \frac{1}{2} \left( 3  \cot{\theta} \; \partial_{\theta} \omega_{(1,1)} +   \partial_{\theta \theta}\omega_{(1,1)} \right)
\nonumber \\
&-&
8 \pi \left(\rho + p\right) r^{2} \; \omega_{(1,1)} 
= 4 \pi \; \frac{\alpha^{2}}{\kappa} \; \frac{e^{\lambda_{(0,0)}}}{\sin{\theta}} \left(1 - \frac{2 m_{(0,0)}}{r} \right)^{1/2}
\left[ \left(\frac{\bar{\rho} - \rho}{r} + \frac{d \rho}{dr} \right) \partial_{\theta} \vartheta_{(1,0)} 
- \left(\bar{\rho}-\rho\right) \partial_{r \theta} \vartheta_{(1,0)} \right]  
\label{Field-eq-final}
\ea
\end{widetext}
Given solutions for
$(\lambda_{(0,0)},m_{(0,0)},\vartheta_{(1,0)},p,\rho)$, we can now solve
the above equation for the CS correction to the metric $\omega_{(1,1)}$.

We can analyze separately the other components of the field
equations. First, we note that all other components of the C-tensor
identically vanish to this order. Moreover, the only non-vanishing
components of the stress-energy tensor are the $(t,t)$, $(r,r)$ and
$(\phi,\phi)$ ones, which are linearly proportional to
$\lambda_{(1,1)}$, $\beta_{(1,1)}$ and $(\gamma_{(1,1)},\chi_{(1,1)})$
respectively.  The Einstein tensor also possesses non-vanishing
components that lead to differential equations for these metric
functions.  However, since the C-tensor vanishes and there are no
metric-independent matter source terms at this order, we can set these
metric perturbations to zero, i.e.,~$(\lambda_{(1,1)},
\beta_{(1,1)},\chi_{(1,1)},\gamma_{(1,1)}) = 0$.

\subsection{Boundary Conditions}

The differential equations presented in the previous subsections must
be complemented by boundary conditions. The exterior solution for the
metric outside a slowly-rotating star in dynamical CS modified gravity
was found in~\cite{Yunes:2009hc}. The line element was found to be
\ba
ds^{2}_{\rm ext} &=& ds^{2}_{\textrm{K}}  +
\frac{5}{4} \frac{\alpha^{2}}{\kappa} \frac{a}{r^{4}}\left( 1 + \frac{12}{7} \frac{M}{r} + \frac{27}{10} \frac{M^{2}}{r^{2}} \right) \sin^{2}{\theta} dt d\phi,
\nonumber \\
\vartheta^{\rm ext} &=&  \frac{5}{8} \alpha \frac{a}{M} \frac{\cos{\theta}}{r^2} \left(1 + \frac{2 M}{r} + \frac{18 M^2}{5 r^2} \right),
\label{theta-BC}
\ea
where $ds^{2}_{\textrm{K}}$ is the slow-rotation limit of the Kerr
line element. The solutions to the differential equations described
in Sec.~\ref{2nd-O-fieldeqs} must be guaranteed to approach the above solution in the limit
$r \to R$, where $R$ is the neutron-star radius.

In terms of the metric functions, the CS correction only
affects the $\omega_{(1,1)}$ term, which becomes, for the exterior metric:
\be
\omega_{(1,1)}^{\rm ext} = - \frac{5}{8} \frac{\alpha^2}{\kappa} \frac{a}{r^{6}} \left( 1 + \frac{12}{7} \frac{M}{r} + \frac{27}{10} \frac{M^{2}}{r^{2}} \right).
\label{omegaBC}
\ee
This relation is derived from the fact that $g_{t \phi}^{(1,0)} = -r^2
\sin^{2}{\theta} \; \omega_{(1,0)}$.  The metric components
corresponding to the Kerr metric have already been given in
Eqs.~\eqref{BC-00}-\eqref{BC-20}. The boundary conditions are then
simply applied by requiring that all metric functions approach their
exterior modified Kerr counterparts at $r = R$.

\section{Numerical Solutions}
\label{Num-Sols}

In this section we present results for the numerical solution to the
modified field equations.  We begin with the standard GR functions and
proceed to solutions for the CS scalar field and the CS correction to
the metric.

\subsection{General Relativistic Solutions}

We first solve numerically the GR equations of neutron star structure
for the FPS EOS.  We use standard numerical techniques to solve these
equations: we reduce the system of
equations~\eqref{Mprime-eq},~\eqref{alphaprime-eq},~\eqref{TOV-eq}
and~\eqref{omega-eq} to first order and then employ a fourth-order,
Runge-Kutta scheme to obtain a numerical solution. The boundary conditions
for the GR solutions are implemented following the scheme of~\cite{1967ApJ...150.1005H,Kalogera:1999nj}, which
exploits the scale-free nature of the differential equations.

\begin{figure}[t]
\epsfig{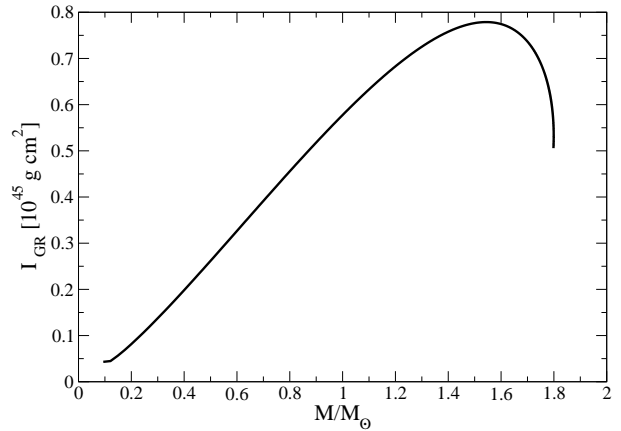}
\caption{\label{fig.IGR} The moment of inertia of neutron stars described
by the FPS EOS, as a function of their masses.}
\end{figure}

We have already shown the mass-radius relation in
Fig.~\ref{fig.eos}. From the $g_{t\phi}$ metric component, we now
compute the moment of inertia via
\be
I \equiv \frac{8 \pi}{3}  \int_{0}^{R} r^{4} \left( \rho + p \right) \sqrt{\frac{g_{rr}}{g_{tt}}} \; \left(1-\frac{\omega(r,\theta)}{\Omega}\right) \; dr\;.
\label{mom-inertia}
\ee
Here $g_{tt}$ and $g_{rr}$ are the $(t,t)$ and $(r,r)$ components of
the metric.  We plot this quantity in Fig.~\ref{fig.IGR}
as a function of neutron-star mass.
%

\subsection{First-Order CS Scalar Field Solution}

In order to solve Eq.~\eqref{scal-evol-first-O}, we will first convert
this partial differential equation to an ordinary one via separation
of variables, i.e.,
\be
\vartheta^{(1,0)}(r,\theta) = \sum_{n} \bar{\vartheta}_{n}(r) P_{n}(\cos\theta),
\label{theta-exp}
\ee
where $P_{n}$ are the Legendre polynomials~\cite{Yunes:2007ss} and we
have omitted the order symbols in $\bar{\vartheta}$ for convenience.

The $\theta$-dependent term in Eq.~\eqref{scal-evol-first-O} is
nothing but
\be
\vartheta^{(1,0)}_{,\theta\theta} + \cot{\theta} \; \vartheta^{(1,0)}_{,\theta} = 
\frac{1}{\sin{\theta}} \partial_{\theta} \left(\sin{\theta} \; \vartheta^{(1,0)}_{,\theta}\right)\;.
\ee
When we evaluate this with respect to Legendre polynomials, we find
\be
\frac{1}{\sin{\theta}} \partial_{\theta} \left(\sin{\theta} \; P_{n}{}_{,\theta}\right) = - n \left( n + 1\right) P_{n}(\cos{\theta})
\label{master-eq-Pn}
\ee
using the master Legendre polynomial equation. With this at hand,
Eq.~\eqref{scal-evol-first-O} becomes
\ba
&& e^{-2 \beta_{(0,0)}} \bar{\vartheta}_{n}''
+  \bar{\vartheta}_{n}' \; e^{-2 \beta_{(0,0)}} \left(\lambda'_{(0,0)} - \beta'_{(0,0)} + \frac{2}{r} \right) 
\nonumber \\
&-&  \frac{n \left(n + 1\right)}{r^{2}} \bar{\vartheta}_{n}   
= \frac{12 \alpha}{r^{3}} e^{-\lambda_{(0,0)}} \frac{\cos{\theta}}{P_{n}} 
\left(m_{(0,0)}  \right.
\nonumber \\
&& \left. \qquad \qquad \qquad  
-  \frac{4 \pi}{3} r^{3} \rho \right) 
\sqrt{1 - \frac{2 m_{(0,0)}}{r}} \; \omega'_{(1,0)}.
\ea
Clearly, all $\theta$ dependence disappears once we choose $n=1$, such
that $P_{1} = \cos{\theta}$ and $\vartheta^{(1,0)} = \bar{\vartheta}
\cos{\theta}$. This ordinary differential equation becomes
\ba
&& \hspace{-0.5cm} e^{-2 \beta_{(0,0)}} \bar{\vartheta}_{1}''
+  \bar{\vartheta}_{1}' \; e^{-2 \beta_{(0,0)}} \left(\lambda'_{(0,0)} - \beta'_{(0,0)} + \frac{2}{r} \right) 
- \frac{2}{r^{2}} \bar{\vartheta}_{1}   
\nonumber \\
&=&  \frac{12 \alpha}{r^{3}} e^{-\lambda_{(0,0)}} \left(m_{(0,0)} 
-  \frac{4 \pi}{3} r^{3} \rho \right) 
\sqrt{1 - \frac{2 m_{(0,0)}}{r}} \; \omega'_{(1,0)},
\nonumber \\
\label{final-Theta-Eq}
\ea
which we can simplify even more using the lower-order TOV equations. The
result is 
\ba
&& \hspace{-0.75cm} \bar{\vartheta}'' + \bar{\vartheta}' \left(1 - \frac{2 m_{(0,0)}}{r}\right)^{-1} \left[ 4 \pi r \left(p - \rho\right) + \frac{2}{r} \left(1 - \frac{m_{(0,0)}}{r} \right)\right]
\nonumber \\
&& \hspace{-0.75cm} - \frac{2 \bar{\vartheta}}{r^{2}} \left(1 - \frac{2 m_{(0,0)}}{r}\right)^{-1} 
= \frac{12 \alpha}{r^{3}} e^{-\lambda_{(0,0)}} \left(m_{(0,0)}  
-  \frac{4 \pi}{3} r^{3} \rho \right) 
\nonumber \\
&\times&
\left(1 - \frac{2 m_{(0,0)}}{r}\right)^{-1/2} \; \omega'_{(1,0)}. 
\label{final-Theta-Eq-2}
\ea
\begin{figure}[t]
\epsfig{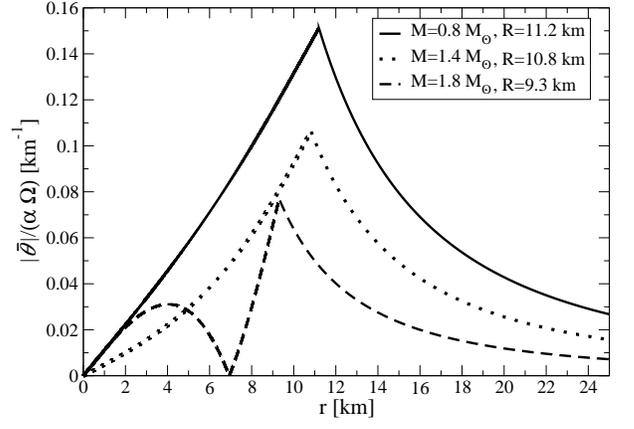}
\caption{\label{fig.CS_Theta} The absolute value of the dipolar
  component of the scalar field $\bar{\vartheta}$ as a function of
  radius, for three stars with different masses and radii.}
\end{figure}

Equation~\eqref{final-Theta-Eq} is an ordinary differential equation
(of dipole, $n=1$, type) that must be solved for $\bar{\vartheta}_{1}$.  The boundary
conditions for this function are the following: at the neutron star
surface, $\bar{\vartheta}_{1} \cos{\theta}$ should equal
Eq.~\eqref{theta-BC}; at the center of the core, $\bar{\vartheta}_{1}'
= 0$, as there can be no CS correction when the spin angular momentum
vanishes. This, of course, presents a numerical challenge, as the
above differential equation is not scale free. We choose to use a
shooting method, where we iterate over a variety of boundary
conditions at the core for $\bar{\vartheta}_{1}$ until the surface
boundary condition is satisfied.

Figure~\ref{fig.CS_Theta} shows the absolute value of the CS scalar
field in units of $\rm{km}^{-1}$ as a function of radius.  In this
plot, we have divided $\bar{\vartheta}$ by $\alpha \Omega$ to give it
units of inverse kilometers, since $\alpha$ has units of squared
length and $\Omega$ of inverse length. The interior solution ($r < R$)
is obtained by numerically solving Eq.~\eqref{final-Theta-Eq-2}. The
exterior solution ($r>R$) is simply that of Eq.~\eqref{theta-BC}. Note
that the CS scalar overall increases in magnitude until the surface of
the star, at which points it decays as the radius cubed. This occurs
because as the radius increases toward the surface, the spin angular
momentum increases until it saturates at $r = R$. It is precisely this
quantity (the spin angular momentum) that the CS scalar field couples to.

The behavior of $\bar{\vartheta}$ is slightly different for the most massive star 
with $M=1.8 M_{\odot}$ relative to the less massive ones. The difference in 
behavior is rooted in the source term of the evolution equation for the scalar field 
[the right-hand side of Eq.~\eqref{final-Theta-Eq-2}]. This source is proportional to 
$\omega_{(1,0)}'$, which by definition is proportional to $dI_{\rm GR}/dr$. Figure~\ref{fig.IGR} 
shows that $dI_{\rm GR}/dr$ changes sign (ie. the slope of the curve changes sign) at around 
$M=1.4 M_{\odot}$: $dI_{\rm GR}/dr>0$ for $M<1.4 M_{\odot}$, while $dI_{\rm GR}/dr<0$ for 
$M>1.4 M_{\odot}$. In turn, this implies that the source term in Eq.~\eqref{final-Theta-Eq-2} 
also changes sign. The change in sign of the source term forces $\bar{\vartheta}$ to start 
negative near the core, but quickly this quantity must switch sign so that the boundary 
condition in Eq.~\eqref{theta-BC} is satisfied at the neutron star surface. This then explains
why $\bar{\theta}$ behaves differently inside the neutron star for stellar configurations
with masses larger or smaller than $1.4 M_{\odot}$.

\subsection{Solution to the Second-Order Field Equations}

As before, we solve Eq.~\eqref{Field-eq-final} numerically. However, it is again
convenient to convert this partial differential equation into an
ordinary one via separation of variables, i.e.,
\be
\omega_{(1,1)}(r,\theta) = \sum_{n} \bar{\omega}_n(r) \left[\frac{1}{\sin{\theta}} \partial_{\theta} P_{n}(\cos{\theta})\right].
\ee

Such a functional dependence for the angular sector is chosen so that
the left-hand side of Eq.~\eqref{Field-eq-final} simplifies to
\ba
&& 3 \cot{\theta} \; \partial_{\theta} \left( \frac{\partial_{\theta} P_{n}}{\sin{\theta}} \right)
+ \partial_{\theta \theta} \left( \frac{\partial_{\theta} P_{n}}{\sin{\theta}} \right)  =
\nonumber \\
&& \qquad - \left(n+2\right)\left(n-1\right) \left( \frac{\partial_{\theta} P_{n}}{\sin{\theta}} \right)
\ea
after repeated application of Eq.~\eqref{master-eq-Pn}. Using this
relation and inserting Eq.~\eqref{theta-exp}, Eq.~\eqref{Field-eq-final}
becomes
\begin{widetext}
\ba
&& \frac{r^{2}}{2} \left(1 - \frac{2 m_{(0,0)}}{r} \right) \; \bar{\omega}_{n}'' \left( \frac{\partial_{\theta} P_{n}}{\sin{\theta}} \right)
+ \left[2 r \left(1 - \frac{2 m_{(0,0)}}{r}\right) - 2 \pi r^{3} \left( \rho + p \right) \right] \; \bar{\omega}_{n}' \left( \frac{\partial_{\theta} P_{n}}{\sin{\theta}} \right)
- \frac{\bar{\omega}_{n}}{2} \left(n+2\right)\left(n-1\right) \left( \frac{\partial_{\theta} P_{n}}{\sin{\theta}} \right)
\nonumber \\
&-&
8 \pi \left(\rho + p\right) r^{2} \; \bar{\omega}_{n} \left( \frac{\partial_{\theta} P_{n}}{\sin{\theta}} \right)
= - 4 \pi \; \frac{\alpha^{2}}{\kappa} \; e^{\lambda_{(0,0)}} \;  \left(1 - \frac{2 m_{(0,0)}}{r} \right)^{1/2}
\left[ \left(\frac{\bar{\rho} - \rho}{r} + \frac{d \rho}{dr} \right) 
\bar{\vartheta}_{1}   
- \left(\bar{\rho}-\rho\right) 
\bar{\vartheta}_{1}'  \right]  
\ea
\end{widetext}
where, in the right-hand side, only the $n=1$ term survives in
the $\vartheta$ expansion. Once more, for the left-hand side of this
equation to be $\theta$ independent, we must choose $n=1$ in the
$\omega^{(1,1)}$ expansion, which then renders $\omega^{(1,1)} = -
\bar{\omega}_{1}$. The above differential equation then simplifies to
\ba
&&  \frac{r^{2}}{2} \left(1 - \frac{2 m_{(0,0)}}{r} \right) \; \bar{\omega}_{1}'' 
+ \left[2 r \left(1 - \frac{2 m_{(0,0)}}{r}\right) 
\right. 
\nonumber \\
&-& \left.
 2 \pi r^{3} \left( \rho + p \right) \right] \; \bar{\omega}_{1}'
-  8 \pi \left(\rho + p\right) r^{2} \; \bar{\omega}_{1} 
= 4 \pi \; \frac{\alpha^{2}}{\kappa} \; e^{\lambda_{(0,0)}}  
\nonumber \\
&\times&
 \left(1 - \frac{2 m_{(0,0)}}{r} \right)^{1/2}
\left[ \left(\frac{\bar{\rho} - \rho}{r} + \frac{d \rho}{dr} \right) 
\bar{\vartheta}_{1}   
- \left(\bar{\rho}-\rho\right) 
\bar{\vartheta}_{1}'  \right].
\nonumber \\
\label{final-omega11-Eq}
\ea
Given that $\bar{\vartheta}_{1}$ is known, the above equation is
simply an ordinary differential equation for $\bar{\omega}_{1}$.

The boundary conditions for Eq.~\eqref{final-omega11-Eq} are the
following: at the neutron star surface, $\bar{\omega}_{1}$ is given by
Eq.~\eqref{omegaBC}; at the center of the core, $\bar{\omega}_{1}' =
0$, as, once more, there can be no CS correction when the spin
vanishes. As before, we employ a shooting method, where we iterate
over boundary conditions at the core for $\bar{\omega}_{1}$ until the
surface boundary condition is satisfied.

\begin{figure}[t]
\epsfig{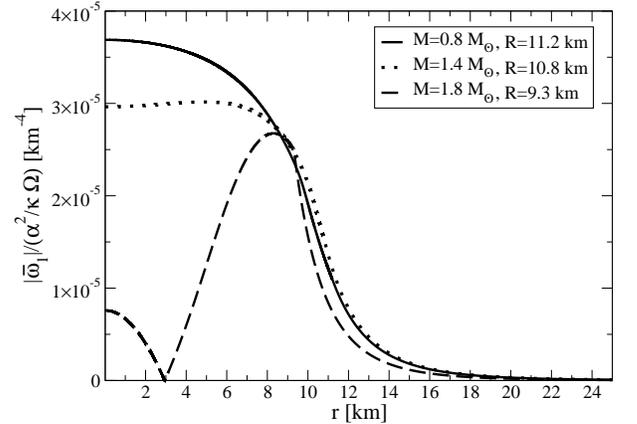}
\caption{\label{fig.CS_Omega} The absolute value of the CS correction
  to the metric coefficient $\bar{\omega}_{1}$ as a function of radius
  inside and outside a neutron star, for stars with three different
  masses and radii.}
\end{figure}

In Fig.~\ref{fig.CS_Omega} we plot the absolute value of the CS
correction to the gravitomagnetic metric component in units of
$\rm{km}^{-4}$ as a function of radius for three stars with different
masses and radii.  The interior solution presents a rather
constant behavior for the two less massive stars, while the exterior
solution always decays with the sixth power of the radius, as
expected.
%

\section{Connection to Observables}
\label{Obs}

In this section, we connect the above CS modifications to observable quantities, in
order to investigate whether neutron star observations can constrain the CS coupling
parameter. For simplicity, let us define $\xi \equiv \alpha^{2}/\kappa$, such 
that $\zeta = \xi/R^{4}$. A constraint on the CS coupling $\alpha$ is thus
a constraint on $\xi$. It is customary to quote constraints on $\xi^{1/4} \propto \alpha^{1/2}$,
as this quantity has units of length.

The mass continuity equation and the equation of hydrostatic equilibrium are CS modified to subleading order, 
i.e.,~${\cal{O}}(2,1)$, because the $\omega$ expansion (see Eq.~\eqref{omega-exp}) cannot have a $\omega_{(0,1)}$ piece
and any correction to the equation of hydrostatic equilibrium is proportional to $\omega^{2} \sim
\omega_{(1,0)}^{2} + 2 \omega_{(1,0)} \omega_{(1,1)} +  2 \omega_{(1,0)} \omega_{(2,0)}$.  This implies
that CS modifications to the mass-radius relation scale as $\omega_{(1,0)} \omega_{(1,1)}  \sim (\xi/R^{4}) (M/R)
\left(R \Omega\right)^{2}$, which is clearly suppressed by a factor of $(R\Omega)^{2} \ll 1$, 
where the factor of $R^{2}$ must be present for dimensional consistency\footnote{This relation can also be obtained by
substituting back into $\omega_{(1,0)} \omega_{(1,1)}$ the surface values required by the boundary conditions.}.
If the mass radius relation where measured to an accuracy of $\delta_{\rm M-R} \sim 10$\%, 
we could then place a constraint of order 
\ba
\xi^{1/4}_{M-R} &\lesssim& 55 \; {\rm{km}} \; \left( \frac{R}{10 \; \rm{km}} \right) \left(\frac{0.025}{R \Omega/c} \right)^{1/2} 
 \nonumber \\
 &\times&
 \left(  \frac{\delta_{\rm M-R}}{0.1} \right)^{1/4}
  \left(\frac{R}{10 \; {\rm{km}}} \frac{1.6 M_{\odot}}{M}\right)^{1/4}\,,
\ea
which derives simply from requiring that $\delta_{\rm M-R} \lesssim  \omega_{(1,1)} \omega_{(1,0)}/(M/R) \sim (\xi/R^{4})
\left(R \Omega\right)^{2}$. We have scaled the previous expression by $R = 10 \; {\rm{km}}$, $M = 1.6 M_{\odot}$ and 
$\Omega/(2 \pi) = 600 \; {\rm{Hz}}$, corresponding to the typical radius and mass and the largest spin frequency of neutron stars 
bursters for which we have measurements~\cite{Ozel:2008kb}. 
Of course, for more slowly rotating pulsars, for example if $\Omega \sim 45 \; {\rm{Hz}}$ 
($\Omega \sim 0.36 \; {\rm{Hz}}$), corresponding to the A (B) component of the double binary pulsar~\cite{Kramer:2009zza}, 
then the bound worsens dramatically: $\xi^{1/4}_{M-R} \lesssim 220 \; {\rm{km}}$ ($\xi^{1/4}_{M-R} \lesssim 2400 \; {\rm{km}}$). 

The weakness of the CS correction to the mass-radius relation of slowly-rotating stars is in fact an advantage. 
For sufficiently slowly-rotating stars, for example, for stars whose spin frequency is $\Omega \ll 10^{3} \; {\rm{Hz}}$, 
corrections of ${\cal{O}}(\epsilon^{2})$ to the mass-radius relation can be ignored. In such cases, the degeneracy between
EOS and CS corrections is broken. That is, given a measurement of the mass-radius relation for a sub-kHz neutron star, 
the CS correction can be ignored, and the mass-radius relation can be used to infer the EOS. 

A stronger constraint arises from the measurement of the moment
of inertia of a neutron star, as is suggested by the modification to the 
gravitomagnetic sector of the metric~ [see
Eq.~\eqref{mom-inertia}]. Since the metric function $\omega$ is CS
modified, we infer that $I$ is also modified. We then
decompose $I = I_{\rm GR} + I_{\rm CS}$, where the latter is given by
Eq.~\eqref{mom-inertia} with the substitution $(1 - \omega/\Omega) \to
-\omega_{(1,1)}/\Omega$ and $(g_{tt},g_{rr})$ given by their GR
values. Since the moment of inertia scales by definition as
$1/\Omega$, the CS modification is first-order in $\epsilon$.
We can then infer that roughly $I \sim I_{\rm
GR} \left(1 + \xi/R^{4} \right)$, and assuming that $I$ 
has been measured to an accuracy $\delta_{I}$ 
(and found to agree with the GR result), we could infer 
the following constraint on $\xi$: 
\be
\xi^{1/4}_{I} \lesssim 5 \; {\rm{km}} \; \left( \frac{R}{10 \; \rm{km}} \right) \left(  \frac{\delta_{I}}{0.1} \right)^{1/4}\;.
\label{I-const}
\ee
Here we have assumed again a fiducial neutron-star radius of $10$ km and the nominal accuracy for the error 
in the measurement of the moment of inertia of $10\%$~\cite{2005ApJ...629..979L,Kramer:2009zza}. 
As expected, this constraint is independent on the rotation rate of the neutron star. 

Another way to estimate the magnitude of the constraint is to consider
directly the $\omega$ metric perturbation.  Since the moment of
inertia depends linearly on $\omega$, a measurement of $I$ to an
accuracy $\delta_{I}$ is effectively a measurement of $\omega$ to the same
accuracy. Given then a measurement $\bar{\omega}$, we can infer 
$\bar{\omega} \left(1 + \delta \right) = \omega_{(1,0)} \left(1 + \omega_{(1,1)}/\omega_{(1,0)} \right)$,
which leads to the constraint $\delta \lesssim \omega_{(1,1)}/\omega_{(1,0)}$, or simply
\be
\xi^{1/4}_{\omega}
\lesssim 5 \; {\rm{km}} \; \left( \frac{R}{10 \; \rm{km}} \right)^{3/4} \left(\frac{M}{1.4 \; M_{\odot}}\right)^{1/4} 
\left(  \frac{\delta_{I}}{0.1} \right)^{1/4},
\ee
where we have here used Eqs.~\eqref{BC-10} and \eqref{omegaBC} to model the mass and radius dependence.
As expected, this constraint is consistent with the one shown in Eq.~\eqref{I-const}.

\begin{figure}[t]
\epsfig{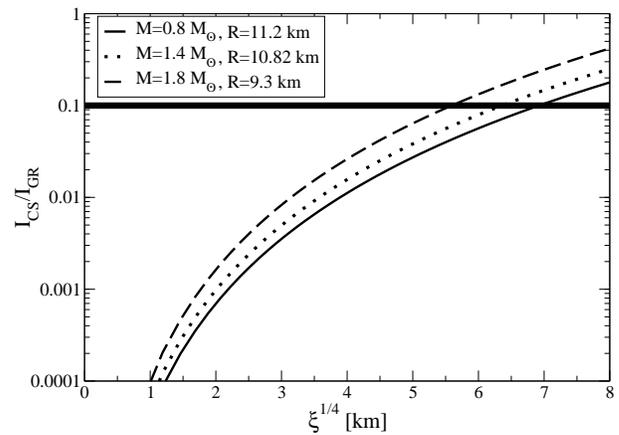}
\caption{\label{fig.CS_I} The ratio between the CS correction to the
  moment of inertia and the GR value as a function of the CS coupling
  parameter $\xi^{1/4}$ in units of kilometers. The thick horizontal
  line corresponds to a measurement of the moment of inertia with an
  accuracy of $10\%$. }
\end{figure}

In order to obtain more specific constraints on the CS coupling, we
solved for the CS correction to the moment of inertia numerically.
In Fig.~\ref{fig.CS_I} we plot the ratio of the CS correction to the
moment of inertia to the GR value as a function of $\xi^{1/4}$
for three stars with different masses and radii (the lowest-mass case is shown for
illustrative reasons, as no astrophysical process is known that leads
to such light neutron stars). We remind the reader that we have previously defined 
$\xi \equiv \alpha^{2}/\kappa$.
A $10\%$ measurement implies a constraint of roughly $\xi^{1/4}
\lesssim (5.5,6.5,7) \; {\rm{km}}$ for stars with final mass $M =
(0.8,1.4,1.8) M_{\odot}$. These constraints are of the same order of
magnitude as the rough estimates presented above. Also notice that the
expansion parameter $\zeta = \xi/R^{4}$ is of order
${\cal{O}}(10^{-1})$ if we saturate these constraints, which justifies
the small-coupling approximation. We should emphasize that, unlike
constraints derivable from measurements of the mass-radius relation,
these constraints are independent of the spin frequency of the neutron star.

The possible bounds that one could achieve with moment of inertia
measurements is to be contrasted from that obtained from observations
of the orbital evolution of the double binary pulsar $\xi_{\rm
  binary}^{1/4} \lesssim 10^{4} \; {\rm{km}}$~\cite{Yunes:2009hc}. The
moment-of-inertia bound will be much stronger because, in the binary
pulsar case, the CS coupling parameter couples to the binary
separation (roughly $10^{5} \; {\rm{km}}$), which is much larger than
the neutron star radius.

\section{Conclusions}
\label{Conclusions}

We have studied the structure of neutron stars in dynamical
CS gravity. We have found that to leading order, only the
gravitomagnetic sector of the metric is CS corrected. This implies
that the mass-radius relation is only modified to second order, while the
moment of inertia is modified to leading order. We have derived the CS
modified equations of stellar structure and we have solved them
numerically. These solutions complete the prescription of the
gravitational field inside and outside of stars.  We estimate that a
measurement of the moment of inertia to $10 \%$ can lead to a
constraint on the theory that is at least three orders of magnitude stronger
than previous constraints.

As is typical with tests of alternative theories with neutron star
observations, the observables are degenerate with the EOS, 
which leads to a degeneracy between alternative theory
modifications and our knowledge of nuclear physics. In the case of CS
gravity, however, the mass-radius relation can be used to extract the
EOS, as neither the mass nor the radius are CS corrected
to leading order. This is simply a consequence of the fact that CS
gravity couples to leading-order to the spin angular momentum and not
to the mass density.

Future observations that lead to a measurement of the moment of
inertia of a neutron star would make the test presented here viable.
The resulting constraint holds the promise of being one of the
strongest achievable with any type of astrophysical observation, at
least in the static sector of the theory. This analysis could then be
complemented by gravitational wave tests, which sample CS corrections
in the radiative sector. 

\acknowledgments

We are grateful to Ramesh Narayan for his insights and assistance with
the numerical solutions, as well as to David Spergel, Frans Pretorius 
and Branson Stephens for useful suggestions and comments. 
Some calculations used the computer algebra
systems MAPLE, in combination with the GRTensorII
package~\cite{grtensor}.  NY acknowledges the support of the NSF grant
PHY-0745779. DP acknowledges the support of the NSF CAREER award
PHY-0746549. This work was supported in part by NSF grant AST-0708640
for F\"O and NSF grant AST-0907890 for AL.

\mbox{}

\appendix

\section{Some Analytic Solutions to the Gravitomagnetic Sector of the Metric}

Equation~\eqref{omega-eq} governs the behavior of the gravitomagnetic
metric perturbation $\omega_{(1,0)}$. Unfortunately, there is no known
closed-form, analytic solution for any realistic EOS. However, for some
particularly simple EOSs, such a solution can in fact be found.

One example is to set $p = - \rho$, corresponding to a
{\emph{gravastar}} configuration~\cite{Mazur:2001fv,Visser:2003ge},
i.e.,~a gravitational vacuum star, whose interior solution is
diffeomorphic to De Sitter spacetime. In this case, the dissipative
function becomes $D = 4/r$, the potential $V=0$, and the right-hand
side of Eq.~\eqref{omega-eq} vanishes. We can then solve for the
gravitomagnetic potential to find
\be
\omega_{(1,0)} = c_{1} + \frac{c_{2}}{r^{3}},
\ee
where $c_{1,2}$ are two constants of integration. Finally, we can also
solve the equation of hydrostatic equilibrium to
obtain $p = c_{3}$, where $c_3$ is a constant of integration.

Perhaps a more realistic configuration is one with vanishing pressure
and constant density $\rho = \rho_{\rm c}$. In this case,
Ref.~\cite{Smith:2007jm} has found that the GR solution is
\be
g^{(1,0)}_{t \phi} = \frac{4 \pi}{3} \rho_{\rm c} R^{2} \left(\vec{r} \times \vec{\Omega}\right)_{\phi} 
\left[ 2 - \frac{6}{5} \left(\frac{r}{R}\right)^{2} \right]
\ee
assuming the fluid is rigidly rotating with angular velocity $\Omega$.  

A third alternative is to allow $\rho + p = (r - 2 m_{(0,0)})/(\pi
r^{3})$. In this case, $D = 0$ and $V=-16/r^{2}$, while the right-hand
side of Eq.~\eqref{omega-eq} becomes equal to $-16 \Omega/r^{2}$.  The
gravitomagnetic component becomes
\be
\omega_{(1,0)} = \Omega + c_{1} r^{p_{1}} + c_{2} r^{p_{2}},
\ee
where $p_{1,2} = 1/2 \pm \sqrt{65}$ and $c_{1,2}$ are constant of
integration.

\bibliographystyle{apsrev}
\bibliography{review}
\end{document}